%% file: pkdd.tex
\documentclass{llncs} 

\usepackage[T1]{fontenc}
\usepackage{latexsym,amsmath,amsfonts,amssymb,braket}
\usepackage{array}
\usepackage{listings}
\usepackage{ifmtarg}
\usepackage{calc}

\usepackage{amsfonts}
\usepackage{amssymb}
\usepackage{graphicx}
\usepackage{caption}
\usepackage{lmodern}

\usepackage{url}
\usepackage[noend]{algpseudocode}
\usepackage[table]{xcolor}
\usepackage[labelfont=bf]{caption}
\usepackage{listings}
\usepackage{courier}
\usepackage{enumitem}
\usepackage{emptypage}
\usepackage{wrapfig,lipsum,booktabs}

\usepackage{graphicx}
\usepackage{subfig}
\usepackage[colorlinks=true,allcolors=blue]{hyperref}

\usepackage{listings}
\usepackage[table]{xcolor}

\definecolor{Gray}{gray}{0.9}

\definecolor{mygray}{rgb}{0.1,0.1,0.1}

\usepackage{tikz}

\begin{document}

\title{An Efficient Algorithm for Mining Frequent Sequence with Constraint Programming}

\author
  {John O.R. Aoga \inst{1} \and
  Tias Guns\inst{2}  \and
   Pierre Schaus\inst{1} }
   
\authorrunning{J.O.R. Aoga, T. Guns and P. Schaus}

\institute
  {UCLouvain, ICTEAM (Belgium) \\
   \email{\{john.aoga,pierre.schaus\}@uclouvain.be}
   \and
   KU Leuven, DTAI Research group (Belgium) \\
   \email{tias.guns@cs.kuleuven.be}
   }

\date{Received: date / Accepted: date}

\maketitle

\begin{abstract}
The main advantage of Constraint Programming (CP) approaches for sequential pattern mining (SPM) is their modularity, which includes the ability to add new constraints (regular expressions, length restrictions, etc).
The current best CP approach for SPM uses a global constraint (module) that computes the projected database and enforces the minimum frequency; it does this with a filtering algorithm similar to the PrefixSpan method.
However, the resulting system is not as scalable as some of the most advanced mining systems like Zaki's cSPADE.
We show how, using techniques from both data mining and CP, one can use a generic constraint solver and yet outperform existing specialized systems.
This is mainly due to two improvements in the module that computes the projected frequencies:
first, computing the projected database can be sped up by pre-computing the positions at which an symbol can become unsupported by a sequence, thereby avoiding to scan the full sequence each time; and second by taking inspiration from the \textit{trailing} used in CP solvers to devise a backtracking-aware data structure that allows fast incremental storing and restoring of the projected database.
Detailed experiments show how this approach outperforms existing CP as well as specialized systems for SPM, and that the gain in efficiency translates directly into increased efficiency for other settings such as mining with regular expressions.

\end{abstract}

\section{Introduction}
\label{intro}
Sequence mining is a widely studied problem concerned with discovering subsequences in a dataset of given sequences, where each (sub) sequence is an ordered list of symbols. It has applications ranging from web usage mining, text mining, biological sequence analysis and human mobility mining~\cite{Mabroukeh:2010:TSP:1824795.1824798}.


In recent years, constraint programming (CP) has been proposed as a general framework for pattern mining~\cite{guns2011itemset,lakdar12,negrevergne2015constraint,kemmar2015global}.
The main benefit of CP-based approaches over dedicated algorithms is that it is \textit{modular}. In a CP framework, a problem is expressed as a set of constraints that the solutions must satisfy. Each such a constraint can be seen as a module, and can range from being as simple as ensuring that a subsequence does not contain a certain symbol at a certain position, up to computing the frequency of a pattern in a database. This modularity allows for flexibility, in that certain constraints such as symbol restrictions, length, regular expressions etc can easily be added and removed to existing problems. Another advantage is that improving the efficiency of one constraint will improve the efficiency of all problems involving this constraint.

However, this increased flexibility can come at a cost. Negrevergne et al.~\cite{negrevergne2015constraint} have shown that a fine-grained modular approach to sequence mining can support any type of constraints, including gap and span constraints and any quality function beyond frequency, but that this is not competitive with state-of-the-art specialized methods. On the other hand, by using a global constraint that computes the pseudo-projection of the sequences in the database similar to PrefixSpan~\cite{pei2001prefixspan}, this overhead can be reduced. Kemmar et al.~\cite{kemmar2015global,kemmar2015prefix} have further extended this work by introducing one constraint (module) for both the pseudo-projection and the frequency pruning. While reasonable performance is obtained, especially for mining under regular expressions, the method does not achieve the same scalability as some of the best specialized systems such as Zaki's cSpade~\cite{zaki2000sequence}.

In this work we focus on the problem of finding patterns in sequences of individual symbols, which is the most commonly used setting in applications such as web log mining and biological sequence analysis.
We improve the \textit{projected frequency} constraint in a modular CP solver by combining ideas from pattern mining as well as from CP. First, we improve the efficiency of computing the projected database and the projected frequency using last-position lists, similar to the LAPIN algorithm~\cite{lapin-spam} but within a PrefixSpan approach. Second, we take into account not just the efficiency of computing the projected database, but also that of storing and restoring the pseudo-projected database during depth-first search.
We use ideas from \textit{trailing} CP solvers to avoid unnecessary copying,
an approach that is applicable to any depth-first pattern mining algorithm.
We show that the resulting system is faster than previous CP-based sequence miners as well as state-of-the-art specialized systems. Furthermore, we show that by improving this one module, these improvements directly translate to other problems using this module, such as regular-expression based sequence mining.


\section{Related works}
\label{sec:rw}
We review specialized methods as well as CP-based approaches. A more thorough review of algorithmic developments is given in~\cite{Mabroukeh:2010:TSP:1824795.1824798}.

\paragraph{\textbf{Specialized methods.}} Introduced by Srikant and Agrawal \cite{agrawal1995mining}, GSP was the first approach to extract sequential patterns from a sequential database. 
Many works have improved on this apriori-based method, typically employing depth-first search.
A seminal work is that of PrefixSpan~\cite{pei2001prefixspan}.
A prefix in this context is a sequential pattern that can only be extended by appending symbols to it. Given a prefix, one can compute the \textit{projected database} of all suffixes of the sequences that have the prefix as a subsequence. This projected database can then be used to compute the frequency of the prefix and of all its 1-extensions (projected frequency). A main innovation in PrefixSpan is the use of a \textit{pseudo-projected} database: instead of copying the entire (projected) database, one only has to maintain pointers to the position in each sequence where the prefix matched.

Alternative methods such as SPADE~\cite{zaki2000sequence} and SPAM~\cite{DBLP:conf/kdd/AyresFGY02} use a vertical representation of the database, having for each symbol a list of sequence identifiers and positions at which that symbol appears.

Yang et al. have shown~\cite{DBLP:conf/dasfaa/YangWK07} that algorithms with either data representation can be improved by precomputing the last position of each symbol in a sequence. This can avoid having to scan the projected database, as often the reason for scanning is to know whether a symbol still appears in the projected sequence.

The standard sequence mining settings have been extended in a number of directions, including user-defined constraints on length or on the gap or span of a sequence such as in the cSPADE algorithm~\cite{zaki2000sequence},
closed patterns~\cite{yan2003clospan} and algorithms that can handle regular expression constraints on the patterns such as SMA~\cite{trasarti2008sequence}.
These constraints are typically hard-coded in the algorithms.

\paragraph{\textbf{CP-based approaches for SPM.}}  CP-based approaches for sequence mining are gaining interest in the CP community. Early work has focused on fixed-length sequences with wildcards~\cite{lakdar12}. More generally, \cite{negrevergne2015constraint} proposed two approaches: a full decomposition of the problem in terms of constraints and an approach using a global constraint to construct the pseudo-projected database similar to PrefixSpan. It uses one such constraint for each sequence.
Kemmar et al \cite{kemmar2015prefix} propose to gather all these constraints into a unique global constraint to reduce the overhead of the multiple constraints. They further showed how the constraint can be modified to take a maximal gap constraint into account~\cite{kemmar2015global}.

\section{Sequential Pattern Mining Background}
\label{sec:pre}
This section introduces the necessary concepts and definitions of sequence mining and constraint programming. 

\subsection{Sequence Mining Background}

Let $I =  \{s_1,\dots,s_N\}$ be a set of $N$ symbols.
In the remaining of the paper when there is no ambiguity a symbol is simply denoted by its identifier $i$ with $i \in \{1,\ldots,N\}$.

\begin{definition}
\textbf{Sequence and sequence database.}  A sequence $s = \langle s_1s_2\dots s_n \rangle$ over $I$ is an ordered list of (potentially repeating) symbols $s_j$, $j \in [1,n]$ with $\#s=n$ the length of the sequence $s$. A set of tuples ($sid$,$s$) where $sid$ is a sequence identifier and $s$ a sequence, is called sequence database ($SDB$).
\end{definition}
\example Table.~\ref{tab:1} shows an example $SDB_1$ over symbols $I = \{A,B,C,D\}$. For the sequence $s=\langle BABC \rangle$: $\#s=4$ and $s_1=B,s_2=A,s_3=B,s_4=C$.

\begin{table}[t]
\centering
\begin{tabular}{llll}
\hline\noalign{\smallskip}
sid $\quad$ & sequence $\quad \quad$ & $lastPosList$ & $lastPosMap$ \\
\noalign{\smallskip}\hline\noalign{\smallskip}
$sid_1$ & $\langle ABCBC \rangle $  & [(C,5),(B,4),(A,1)] & \{A$\rightarrow$1, B$\rightarrow$4, C$\rightarrow$5,D$\rightarrow$0\}  \\
$sid_2$ & $\langle BABC \rangle$   & [(C,4),(B,3),(A,2)] & \{A$\rightarrow$2, B$\rightarrow$3, C$\rightarrow$4,D$\rightarrow$0\} \\
$sid_3$ & $\langle AB \rangle$  & [(B,2),(A,1)] & \{A$\rightarrow$1, B$\rightarrow$2, C$\rightarrow$0,D$\rightarrow$0\} \\
$sid_4$ & $\langle BCD \rangle$   & [(D,3),(C,2),(B,1)]& \{A$\rightarrow$0, B$\rightarrow$1, C$\rightarrow$2,D$\rightarrow$3\} \\
\noalign{\smallskip}\hline
\end{tabular}
\smallskip
\caption{A sequence database $SDB_1$ and list of last positions.}
\label{tab:1}  \label{tab:2} \label{tab:3}
\textbf{ 1) SDB, 2) lastPosList, 3) lastPosMap}
\end{table}

\begin{definition}
\textbf {Sub-sequence ($\preceq$), super-sequence.} A sequence $\alpha = \langle \alpha_1\dots\alpha_m\rangle$ is called a sub-sequence of $s = \langle s_1s_2\dots s_n \rangle$ and $s$ is a super-sequence of $\alpha$ iff (i) $m\le n$ and (ii) for all $i\in [1,m]$ there exist integers $j_i $ s.t. $1\le j_1\le\dots\le j_m\le n$, such that $\alpha_i = s_{j_i}$. 
\end{definition}

\example For instance $\langle BD \rangle$ is a sub-sequence of $\langle BCCD \rangle$, and inversely $\langle BCCD \rangle$ is the super-sequence of $\langle BD\rangle$ : $\langle BD\rangle\preceq\langle BCCD\rangle$.

\begin{definition}
\textbf{Cover, Support, Pattern, Frequent Pattern.} The cover of sequence $p$ in $SDB$, denoted by $cover_{SDB}(p)$, is the subset of sequences in $SDB$ that are a super-sequence of $p$, i.e. $cover_{SDB}(p) = \{ (sid,s) \in SDB \,|\, p\preceq s\}$. The support of $p$ in $SDB$, denoted by $sup_{SDB}(p)$, is the number of super-sequencs of $p$ in $SDB$: $sup_{SDB}(p) = \#cover_{SDB}(p)$.
Any sequence $p$ over symbols in $I$ can be a pattern, and we call a pattern frequent iff $sup_{SDB}(p) \ge \theta$, where $\theta$ is a given minimum support threshold.
\end{definition} 

\example Assume $p=\langle BC\rangle$ and $\theta=2$,  $cover_{SDB_1}(p) = \allowbreak \{ (sid_1,\langle ABCBC \rangle),\allowbreak (sid_2,\langle BABC \rangle),\allowbreak (sid_4,\langle BCD \rangle)\}$  and hence $sup_{SDB_1}(p) =3$.  
Hence, $p$ is a frequent pattern for that given threshold. 

The sequential pattern mining (SPM) problem, first introduced by Agrawal and Srikant \cite{agrawal1995mining}, is the following:

\begin{definition}\textbf{Sequential Pattern Mining (SPM).} Given an minimum support threshold $\theta$ and a sequence database $SDB$, the SPM problem is to find all patterns $p$ such that $sup_{SDB}(p) \ge \theta$.
\end{definition}

%

Our method uses the idea of a \textit{prefix} and \textit{prefix-projected} database for enumerating the frequent patterns. These concepts were first introduced in the seminal paper that presented the \textit{PrefixSpan} algorithm~\cite{pei2001prefixspan}.

\begin{definition}\textbf{Prefix, prefix-projected database} 
Let $\alpha =\langle\alpha_1\dots\alpha_m\rangle$ be a pattern. If a sequence $\beta =\langle\beta_{1}\dots\beta_n\rangle$ is a super-sequence of $\alpha$: $\alpha \preceq \beta$, then the \textit{prefix} of $\alpha$ in $\beta$ is the smallest prefix of $\beta$ that is still a super-sequence of $\alpha$: $\langle \beta_1\dots\beta_j\rangle$ s.t. $\alpha \preceq \langle \beta_1\dots\beta_j\rangle$ and $\nexists j' < j: \alpha \preceq \langle \beta_1\dots\beta_{j'}\rangle$. The sequence $\langle \beta_{j+1}\dots\beta_n\rangle$ is called the suffix and is obtained by \textit{projecting} the prefix away. A prefix-projected database of a pattern $\alpha$, denoted by $SDB|_\alpha$, is the set of prefix-projections of all sequences in $SDB$ that are a super-sequence of $\alpha$.
\end{definition}

\example In $SDB_1$, assume $\alpha=\langle A\rangle$, then $SDB_1|_\alpha=\allowbreak \{ (sid_1,\langle BCBC \rangle),\allowbreak (sid_2,\langle BC \rangle),\allowbreak (sid_3,\langle B \rangle)\}$.

We say that the \textit{prefix-projected frequency} of the symbols $I$ in a prefix-projected database is the number of sequences in which these symbols appear. For $SDB_1|_{\langle A \rangle}$ the prefix-projected frequencies are $A: 0, B: 3, C: 2, D: 0$.

The PrefixSpan algorithm solves the SPM problem by starting from the empty pattern and extending this pattern using depth-first search.
At each step it extends a pattern by a symbol and projects the database accordingly. 
The appended symbol is removed on backtrack. It hence grows the pattern incrementally, which is why it is called a pattern-growth method. A frequent pattern in the projected database is also frequent in the original database.

There are two important considerations for the efficiency of the method. The first is that one does not have to consider during search any symbol that is not frequent in the prefix-projected database. The second is that of \textit{pseudo-projection}: to store the prefix-projected database during the depth-first search, it is not necessary to store (and later restore) an entire copy of the projected database. Instead, one only has to store for each sequence the pointer to the position $j$ that marks the end of the prefix in that sequence (remember, the \textit{prefix} of $\alpha$ in $\beta$ is the smallest prefix $\langle \beta_1\dots\beta_j\rangle \succeq \alpha$). 

\example The projected database $SDB_1|_\alpha=\allowbreak \{ (sid_1,\langle BCBC \rangle),\allowbreak (sid_2,\langle BC \rangle),\allowbreak (sid_3,\langle B \rangle)\}$ can be represented as a pseudo-projected database as follows: $\{ (sid_1,2),\allowbreak (sid_2,3),\allowbreak (sid_3,2)\}$.

\subsection{Constraint Programming Background}
\label{sec:cp} \label{sec:trailing}

CP is a powerful declarative paradigm to solve combinatorial satisfaction and
optimization problems (see, e.g., \cite{cphandbook}). 
A CP problem  $(V,D,C)$ is defined by a set of variables $V$ with their
respective domain $D$ (the values that can be assigned to a variable), and a
set of constraints $C$ on these variables. A solution of a CP problem is an
assignment of the variables to a value from its domain, such that all constraints are satisfied.

At its core, CP solvers are depth-first search algorithms that iterate between \textit{searching} over unassigned variables and \textit{propagating} constraints. Propagation is the act of letting the constraints in $C$ remove unfeasible values from the domains of its variables. This is repeated until \textit{fixed-point}, that is, no more constraint can remove any unfeasible values. Then, a search exploration step is taken by choosing an unassigned variable and assigning it to a value from its current domain, after which propagation is executed again.

\example Let there be 2 variables $x,y$ with domains $D(x) = \{1,2,3\}, D(y) = \{3,4,5\}$. Then constraint $x + y \geq 5$ can derive during propagation that $1 \notin D(x)$ because the lowest value $y$ can take is $3$ and hence $x \geq 5 - \min(D(y)) \geq 5-3 \geq 2$.

\paragraph{Constraints and global constraints}
Many different constraints and their propagation algorithms have been investigated in the CP community. This includes logical and arithmetic ones like the above, up to constraints for enforcing regular expressions or graph theoretic properties. A constraint that enforces some non-trivial or application-dependent property is often called a \textit{global constraint}. For example, \cite{negrevergne2015constraint} introduced a global constraint for the pseudo-projection of a single sequence, and \cite{kemmar2015global} for the entire projected frequency subproblem.

\paragraph{State restoration in CP}
In any depth-first solver, there must be some mechanism to store and restore some \textit{state}, such that computations can be performed incrementally and intermediate values can be stored.
In most of the CP solvers\footnote{One notable exception is the Gecode copy-based solver.} a general mechanism, called  \emph{trailing} is used for storing and restoring the state (on backtrack)~\cite{schulte2006finite}.
Externally, the CP solvers typically expose some "reversible" objects whose values are automatically stored and restored on the trail when they change. The most important example are the domains of CP variables. Hence, for a variable the domain modifications (\emph{assign}, \emph{removeValue}) are automatically reversible operations.
A CP solver also exposes reversible version of primitive types such as integers and sets for use within constraint propagators. They are typically used to store incremental computations.
CP solvers consist of an efficient implementation of the DFS backtracking algorithm, as well as many constraints that can be called by the fix-point algorithm. The modularity of constraint solvers stems from this ability to add any set of constraints to the fix-point algorithm.

\section{Global constraints for projected frequency}
We first introduce the basic CP model of frequent sequence mining introduced in~\cite{negrevergne2015constraint} and extended in~\cite{kemmar2015prefix}. Then, we present how we improve the computation of the pseudo-projection, followed by the projected frequency counting and pruning.

\subsection{Existing methods~\cite{negrevergne2015constraint,kemmar2015prefix}}
As explained before, a constraint model consists of variables, domains and constraints. The CP model will be such that a single solution corresponds to a frequent sequence, meaning that all sequences can be extracted by enumerating all solutions.

Let $L$ be an upper bound on the pattern length, e.g. the length of the longest sequence in the database.
The variables used to represent the unknown pattern $P$ is modeled as an array of $L$ integer variables $P=[P_1,P_2,\dots,P_L]$.
Each variable has an initial domain $\{0,\ldots,N\}$, corresponding to all possible symbols identifiers and augmented with an additional identifier 0. The symbol with identifier 0 represents $\epsilon$, the empty symbol. It will be used to denote the end of the sequence in $P$, using a trailing suffix of such $0$'s.

\begin{definition}
A CP model over $P$ represents the frequent sequence mining problem with threshold $\theta$, iff the following three conditions are satisfied by every valid assignment to $P$:
\begin{enumerate}
\item $P_1 \neq 0$
\item $\forall i \in \{2,\ldots,L-1\}: P_i = 0  \Rightarrow P_{i+1} = 0$
\item $\#\{(sid,s) \in SDB \,\, \langle P_1 \dots P_j\rangle \preceq s\} \ge \theta$, $j = \max(\{i \in \{1\ldots L\} | P_i \neq 0\})$. 
\end{enumerate}
\end{definition}

The first requirement states that the sequence may not start with the empty symbol, e.g. no empty sequence. The second requirement enforces that the pattern is in a canonical form such that after the empty symbol, all other symbols are the empty symbol too. Hence, a sequence of length $l < L$ is represented by $l$ non-zero symbols, followed by $L-l$ zero symbols. The last requirement states that the frequency of the non-zero part of the pattern must be above the threshold $\theta$.

\paragraph{Prefix projection global constraint}
\label{ssec:gpppm}

Initial work~\cite{negrevergne2015constraint} proposed to decompose these three conditions into separate constraints, including a dedicated global constraint for the inclusion relation $\langle P_1 \dots P_j\rangle \preceq s$ for each sequence separately. It used the pseudo-projection technique of PrefixSpan for this, with the projected frequency enforced on each symbol in separate constraints.

Kemmar et al.~\cite{kemmar2015prefix} extended this idea by encapsulating the filtering of all three conditions into one single (global) constraint called 
\texttt{PrefixProjection}. It also uses the pseudo-projection idea of PrefixSpan, but over the entire database.
The propagation algorithm for this constraint, as executed when the next unassigned variable $P_i$ is assigned during search, is given in Listing \ref{code:filterprefixprojection}.

\begin{figure}[t]
\begin{lstlisting}[caption={PrefixProjection(SDB,P,i,$\theta$)}, label=code:filterprefixprojection,escapechar=|]
// pre: variables $\langle P_1,\ldots,P_{i}\rangle$ are bound, $SDB$ is given
//      $P_i$ is the new instantiated variable since previous call.
if ($P_i == 0$) {
  foreach (j $\in \{i+1,\ldots,L\}$) { $P_j.assign(0)$ }
} else if (i $\ge$ 2) {
  projFreqs = ProjectAndGetFreqs($SDB, P_{i}, \theta$) |\label{line:project}|
  foreach (j $\in \{i+1,\ldots,L\}$) |\label{line:filterloop}|
    foreach ($a \in D(P_j)$)
      if ($a \neq 0$ and projFreqs[a] < $\theta$) { $P_j.removeValue(a)$ }     
}
\end{lstlisting}
\end{figure}

An initial assumption is that the database $SDB$ does not contain any infrequent symbols, which is a simple preprocessing step.
The code is divided in three parts: (i) if $P_i$ is assigned to $0$ the remaining $P_k$ with $k > i$ is assigned to $0$; else (ii) from the second position onwards (remember that the first position can take any symbol and be guaranteed to be frequent as every symbol is known to be frequent),  the projected database and the projected frequency of each symbol is computed; and (iii) all symbols that have a projected frequency below the threshold are removed from the domain of the subsequent pattern variables.    

The algorithm for computing the (pseudo) projected database and the projected frequencies of the symbols is given in Listing \ref{code:projsdb}. It operates as follows with $a$ the new symbol appended to the prefix of assigned variables since previous call. The first loop at line \ref{line:firstloop} attempts to discover for each sequence $s$ in the projected database if it can be a sub-sequence of the extended prefix. If yes, this sequence is added to the next projected database at line \ref{line:addpdb}. 
The second loop at line \ref{line:frequencycomp} computes the frequency of each symbol occurring in the projected database but counting it at most once per sequence.


\begin{figure}[t]
\begin{lstlisting}[caption={ProjectAndGetFreqs(SDB,a,$\theta$)}, label=code:projsdb,escapechar=|]
$PSDB_i = \emptyset $
foreach (sid,start) $\in$ $PSDB_{i-1}$ { |\label{line:firstloop}|
  s = SDB[sid]; $pos$ = start
  while ($pos < \#$s and $a \neq $s$[pos]$) { $pos = pos + 1$ } |\label{line:proj}|
  if ($pos < \#$s) { $PSDB_i = PSDB_i \cup \{(sid,pos)\}$ }  |\label{line:addpdb}|
}
projFreqs[a]=0 $\forall a \in \{1,\ldots,N\} $
if ($\#PSDB_i \geq \theta$) {
  foreach (sid,start) $\in PSDB_i$ { |\label{line:frequencycomp}|
    s = SDB[sid]; existsSymbol[b] = false $\forall b \in \{1,\ldots,N\} $
    foreach (i $\in \{$start$,\ldots,\#$s$\}$) {
      if (!existsSymbol[s[i]]) {
        projFreqs[s[i]] = projFreqs[s[i]]+1
        existsSymbol[s[i]] = true
      }
} } }
return projFreqs
\end{lstlisting}
\end{figure}

\subsection{Improving propagation}
Although being the state-of-art approach for solving SPM with CP, the filtering algorithm of Kemmar et al \cite{kemmar2015global} presents room for improvement. We identify four weaknesses and propose solutions to them.

\noindent\textbf{Weakness 1.} Databases with long sequences will have a large upper-bound $L$. For such databases, removing infrequent symbols from all remaining pattern variables $P$ in the loop defined at line \ref{line:filterloop} of Listing~\ref{code:filterprefixprojection} can take time. This is not only the case for doing the action, but also for restoring the domains on backtracking. On the other hand, only the next pattern variable $P_{i+1}$ will be considered during search, and in most cases a pattern will never actually be of length $L$, so all subsequent domain changes are unnecessary.
This weakness is a peculiarity of using a fixed-length array $P$ to represent a variable-length sequence. Mining algorithms typically have a variable length representation of the pattern, and hence only look one position ahead. In our propagator we only remove values from the domain of $P_{i+1}$.

\noindent\textbf{Weakness 2.} When computing the projected frequencies of the symbols, one has to scan each sequence from its current pseudo-projection pointer $start$ till the end of the sequence. This can be time consuming in case of many repetitions of only a few symbols for example.
Thanks to the $lastPosList$ defined next, it is possible to visit only the last position of each symbol occurring after $start$. This idea was first introduced in \cite{DBLP:conf/dasfaa/YangWK07} and exploited in the LAPIN family of algorithms.

\begin{definition} \textbf{(Last position list).}
\label{ssec:nlpit}
For a current sequence $s$, $lastPosList$ is a sequence of pairs $(symbol,pos)$ giving for each $symbol$ that occurs in $s$ its last position: $pos = \max \{p \le \#s: s[p]=symbol \}$. The sequence is of length $m$, the number of distinct symbols in $s$. This sequence is decreasing according to positions: $lastPosList[i].pos > lastPosList[i+1].pos$ $\forall i \in \{1,\ldots,m-1\}$.
\end{definition}

\example Table.~\ref{tab:3} shows the $lastPosList$ sequences for $SDB_1$. 
We consider the sequence with $sid_1$ and a prefix $\langle A \rangle$.
The computation of the frequencies starts at position $2$, remaining suffix is $\langle BCBC\rangle$. Instead of visiting all the 4 positions of this suffix, only the last two can be visited thanks to the information contained in $lastPosList[sid_1]$. Indeed according to $lastPosList[sid_1][1]$ the maximum last position is $5$ (corresponding to the last $C$). Then according to $lastPosList[sid_1][2]$ the second maximum last position is $4$ (corresponding to the last position of symbol $B$). The third maximum last position is $1$ for symbol $A$. Since this position is smaller than $2$ (our initial start), we can stop.

\noindent\textbf{Weakness 3.} Related to weakness 2, line \ref{line:proj} in Listing \ref{code:projsdb} finds the new position ($pos_s$) of $a$ in $SDB[sid]$. This code is executed even if the new symbol no longer appears in that sequence. Currently, the code has to loop over the entire sequence until it reaches the end before discovering this.

Assume that the current position in the sequence $s$ is already larger than the position of the last occurrence of $a$. Then we immediately know this sequence cannot be part of the projected database. To verify this in $O(1)$ time, we use a $lastPosMap$ as follows:

\begin{definition} \textbf{(Last position map of symbols).}\label{ssec:lpit} 
For a given sequence $s$ with id $sid$, $lastPosMap[sid]$ is a map such that $lastPosMap[sid][i]$ is the last position of symbol $i$ in the sequence $s$.  In case the symbol $i$ is not present: $lastPosMap[sid][i]=0$ (positions are assumed to start at index 1).
\end{definition}

\example Table~\ref{tab:2} shows the $lastPosMap$ arrays next to $SDB_1$. For instance for $sid_2$ the last position of symbol $C$ is $4$.

\noindent\textbf{Weakness 4.} Listing \ref{code:projsdb} creates a new set $PSDB_i$ to represent the projected database. This projected database is computed many times during the search, namely at least once in each node of the search tree (more if there are other constraints in the fixPoint set). This is a source of inefficiency for garbage collected languages such as Java but also for C since it induces many "slow" system calls such as free and malloc leading to fragmentation of the memory.
We propose to store and restore the pseudo-projected databases with reversible vectors making use of CP trailing techniques.
The idea is to use one and the same array throughout the search in the propagator, and only maintain the relevant start/stop position during search. Each call to propagate will read from the previous start to stop position, and write after the previous stop position plus store the new start/stop position. The projected data-bases are thus \emph{stacked} in the array along a branch of the search tree. We implement the pseudo-projected database with two reversible vectors: $sids$ and $poss$ respectively for the sequence ids and the current position in the corresponding sequences.  The position $\phi$ is the start entry (in $sids$ and $poss$) of the current projected database,
and $\varphi$ is the size of the projected database. We thus have the current projected database contained in sub-arrays $sids[\phi,\ldots,\phi+\varphi-1]$ and $poss[\phi,\ldots,\phi+\varphi-1]$.
In order to make the projected database reversible, $\phi$ and $\varphi$ are reversible integers.
That is on backtrack to an ancestor node those integers retrieve their previous value and entries of $sids$ and $poss$ starting from $\phi$ can be reused.

\begin{figure}[t]
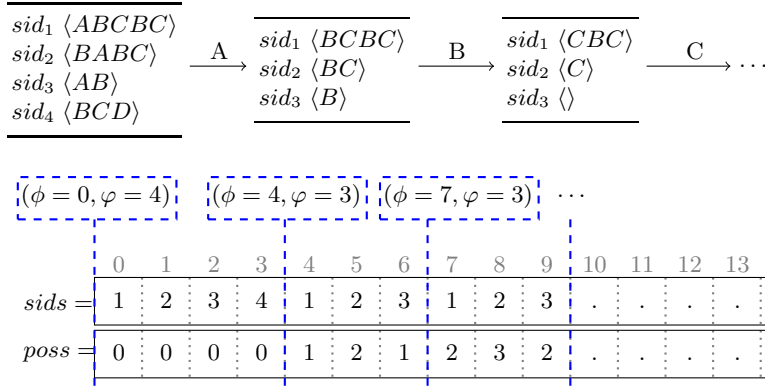

\centering
  \include{fig1}
\caption{Reversible vectors technique}
\label{fig:1} 
\end{figure}

\example Figure~\ref{fig:1} is an example using $SDB_1$. Initially all the sequences are present $\varphi=4$ and position is initialized $\phi=0$. The $A$-projected database contains sequence $1,2,3$ at positions $1,2,1$ with $\phi = 4$ and $\varphi=3$.

\subsubsection{\textbf{Prefix Projection Incremental Counting propagator (\textsc{PPIC}).}}
Putting all the solutions to the identified weaknesses together, we list the code of the main function of our propagator's in Listing \ref{code:ourprojsdb}.

The main loop at line \ref{line:mainloop} iterates over the previous \textit{(parent)} projected database.
In case the sequence at index $i$ in the projected database contains the new symbol at a subsequent position larger or equal to $start$, the matching position is searched and added to the new projected database (at index $j$ of reversible vectors $sids$ and $poss$) at line \ref{line:addsupport}. 
Then the contribution of the sequence to the projected frequencies is computed in the loop at line \ref{line:countfreq}. Only the entries in the \texttt{lastPosList} with position larger than current $pos$ are considered (recall that his list is decreasing according to positions).
Finally line \ref{line:trailphi} updates the reversible integers $\phi$ and $\varphi$ to reflect the newly computed projected data-base. 
Based on these projected frequencies a filtering similar to the one of Listing \ref{code:filterprefixprojection} is achieved except that only the domain of the next variable $D(P_{i+1})$ is filtered according to the solution to Weakness 1.
				
\begin{figure}[t]
\begin{lstlisting}[caption={ProjectAndGetFreqs(SDB,$a,\theta$,$sids$,$poss$,$\phi$,$\varphi$)},escapechar=|, label=code:ourprojsdb]
projFreqs[b]=0 $\forall b \in \{1,\ldots,N\} $
$i = \phi$; $j = \phi + \varphi$; $sup = 0$
while ($i < \phi + \varphi$) { |\label{line:mainloop}|
  $sid = sids[i]$; $pos = poss[i]$; s = SDB[$sid$]
  if (lastPosMap$[sid][a] - 1 \ge start$) {  |\label{line:okseq}|
    //find the next position of $a$ in s
    while ($pos < \#$s and $a \neq $s$[pos]$) { $pos = pos + 1$ } |\label{line:findnext}|
    // update projected database
    $sids[j] = sid$; $poss[j] = pos + 1$; $j = j + 1$; $sup = sup + 1$ |\label{line:addsupport}|
    // recompute projected frequencies
    foreach ($(symbol,pos_x)$ in lastPosList$[sid]$) { |\label{line:countfreq}|
        if ($pos_x \leq pos$) { break }
        projFreqs[symbol] = projFreqs[symbol] + 1
  } }
  $i = i + 1$
}
$\phi = \phi + \varphi$; $\varphi = sup$ |\label{line:trailphi}|
return projFreqs
\end{lstlisting}
\end{figure}

\subsubsection{\textbf{Prefix Projection Decreasing Counting propagator (\textsc{PPDC}).}}

The key idea of this approach is not to count the projected frequencies from scratch, but rather to \textit{decrement} them.
More specifically, when scanning the position of the current symbol at line \ref{line:findnext}, if
$pos$ happens to be the a symbol last position (\texttt{pos==lastPosMap[sid][s[pos]]}) then \texttt{projFreqs[s[pos]]} is decremented. This requires \texttt{projFreqs} to be an array of reversible integers.
With this strategy the loop at line \ref{line:countfreq} disappears, but in case the current sequence is not added to the projected data-base, the frequencies of all its last symbols occurring after $pos$ must also be decremented. This can be done by adding an \textbf{\texttt{else}} block to the \textbf{\texttt{if}} defined at line \ref{line:okseq} that will iterate over the \texttt{lastPosList} and decrement the symbol frequencies.

\example Assume $SDB_1$. The initial projected frequency array is \texttt{projFreqs=} \texttt{[A:3,B:4,C:3,D:1]}. 
Consider now the $A$-projected data-base illustrated on Fig.~\ref{fig:1}. The projected frequency array becomes \texttt{projFreqs=}\texttt{[A:0,B:3,C:2,D:0]}. The entry at \texttt{A} is decremented three times as $pos$ moved beyond its $lastPos$ for each of the sequences $sid_1$, $sid_2$ and $sid_3$. Since $sid_4$ is removed from the projected data-base, the frequency of all its last symbols occurring after $pos$ is also decremented, that is for entries $B$, $C$ and $D$.

\textbf{PP-mixed.} Both PPID and PPDC approaches can be of interest depending on the number of removed sequences in the projected data-base. If the number of sequences removed is large then PPIC is preferable.
On the other hand is only a few sequences are removed then PPDC can be more interesting.
Inspired from the \emph{reset} idea of \cite{perez2014improving} the PP-mixed approach dynamically chooses the best strategy: if $projFreqs_{SDB}(a) < \#PSDB_i/2$ (i.e., more than half of sequences will be removed) then PPIC is used otherwise PPDC.

\subsection{Constraints of SPM} We implemented common constraints such as minimum and maximum pattern size, symbol inclusion/exclusion, and regular expression constraints. 
Time constraints (maxgap, mingap, maxspan,etc) are outside the scope of this work: they change the definition of what a valid prefix is, and hence require changing the propagator (as in \cite{kemmar2015global}).

\section{Experiments}
\label{sec:exp}
In this section, we report our experimental results on the performance of our approaches with six real-life datasets\footnote{\scriptsize\url{http://www.philippe-fournier-viger.com/spmf/}} and one synthetic (data200k~\cite{trasarti2008sequence}) with various characteristics shown in Table.~\ref{tab:4}.
Sparsity, representing the average of the number of symbols that appear in each sequence, is a good indicator of how sparse or dense a dataset is. 

Our work is implemented in Scala in OscaR solver \cite{oscar} and run under JVM with maximum memory set to 8GB. All our software, datasets and results are available online as open source in order to make this research reproducible (\small\url{http://sites.uclouvain.be/cp4dm/spm/}).

We used a machine with a 2.7Hz Intel core i5 processor and 8GB of RAM with Linux 3.19.0-32-generic 64 bits distribution Mint 17.3. Execution time limit is set to 3600 seconds (1 hour).
Our proposals are compared, first, with CPSM\footnote{\scriptsize\url{https://dtai.cs.kuleuven.be/CP4IM/cpsm/}}\cite{negrevergne2015constraint} and  \textsc{Gap-Seq}\footnote{\scriptsize\url{https://sites.google.com/site/cp4spm/}}\cite{kemmar2015global}, the recently CP-based approaches including Gap constraint and the previous version of \textsc{Gap-Seq}, \textsc{PP}\footnote{\scriptsize\url{https://sites.google.com/site/prefixprojection4cp/}}\cite{kemmar2015prefix} without Gap but with regular expression constraint. 
Second,  we made comparison with \textsc{cSpade}\footnote{\scriptsize\url{http://www.cs.rpi.edu/~zaki/www-new/pmwiki.php/Software}}\cite{zaki2000sequence}, PrefixSpan~\cite{pei2001prefixspan}\footnote{\scriptsize\url{http://goo.gl/goqHPL}} 
and SPMF\footnote{\scriptsize\url{http://www.philippe-fournier-viger.com/spmf/index.php?link=download.php}}.

\begin{table}[t]
\caption{Dataset Features. Sparsity is equal to ($ \frac{1}{\#SDB}\times\sum \frac{\#s}{\#I_{/s}}$)}
\label{tab:4}       
\begin{tabular}{llllllll}
\hline\noalign{\smallskip}
 SDB & $\#SDB$  & $N$  & $avg(\#s)$& $avg(\#I_{/s})$  & $max(\#s)$  & sparsity & description    \\
\noalign{\smallskip}\hline\noalign{\smallskip}
BIBLE & 36369 &  13905 & 21.64 & 17.85 & 100 & 1.2 & text  \\
FIFA & 20450 & 2990 & 36.24 & 34.74 & 100 &  1.2 & web click stream\\
Kosarak & 69999  & 21144 & 7.98 & 7.98 & 796  & 1.0 &   web click stream\\
Leviathan & 5834  & 9025 & 33.81 & 26.34 & 100  & 1.3 & text\\
PubMed & 17237  & 19931 & 29.56 & 24.82 & 198  & 1.2 &  bio-medical text\\
data200k & 200000  & 26 & 50.25 & 18.25 & 86  & 2.8 & synthetic data\\
protein & 103120 &  25 & 482.25 & 19.93 & 600  & 24.2 & protein sequences\\
\noalign{\smallskip}\hline
\end{tabular}
\end{table} 

\begin{figure*}[!h]
  \includegraphics[width=\textwidth]{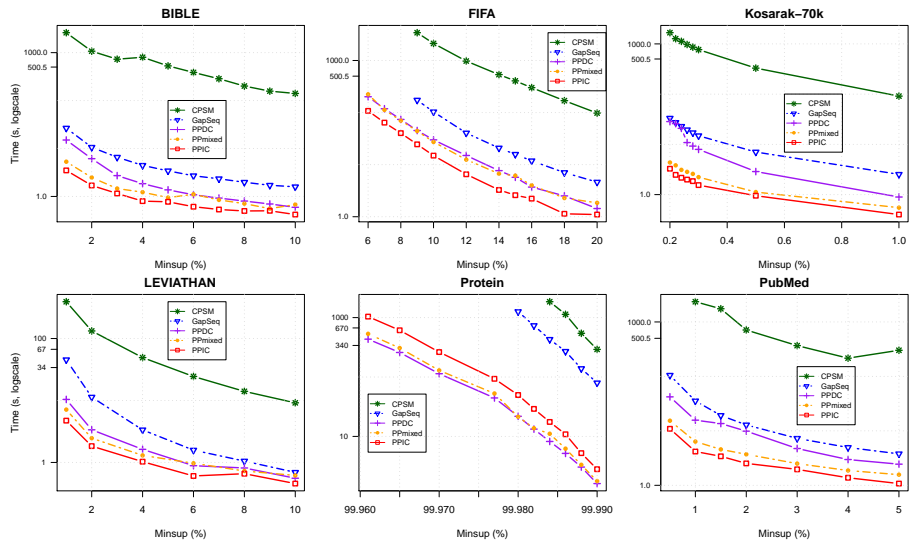}
\caption{CPU times for PPIC, PPDC, \textsc{PPmixed} and \textsc{Gap-Seq} for several minsup (missing points indicate a timeout)}
\label{fig:3}       
\end{figure*}

\paragraph{\textbf{PPIC vs PPDC vs PPmixed}.} The CPU time of PPIC, PPDC and \textsc{PPmixed} models are shown in Fig.~\ref{fig:3}. 
PPIC is more efficient than PPDC in 80\% of datasets. 
This is essentially because in many cases at the beginning of mining, there are many unsupported sequences for which the symbol counters must be decremented (compared to not having to increase the counters in PPIC).
For instance with BIBLE SDB and $minsup = 10\%$ PPDC need to see 21,979,585 symbols to be complete while only 15,916,652 is needed for PPIC. 
Unsurprisingly, \textsc{PPmixed} is between these approaches.

\paragraph{\textbf{Our proposals vs \textsc{Gap-Seq}  (CP method)}.} Fig.~\ref{fig:3} confirms CPSM is outperformed by \textsc{Gap-Seq} and shows that \textsc{Gap-Seq} improves on PP even without gap. We can clearly notice our approaches outperform \textsc{Gap-Seq} (and hence PP) in all cases.
In the case of FIFA SDB, \textsc{Gap-Seq} reach time limit when $minsup \le 9\%$.
PPIC is very effective in large and dense datasets regarding of CPU-times.

\begin{figure*}[t]
  \includegraphics[width=\textwidth]{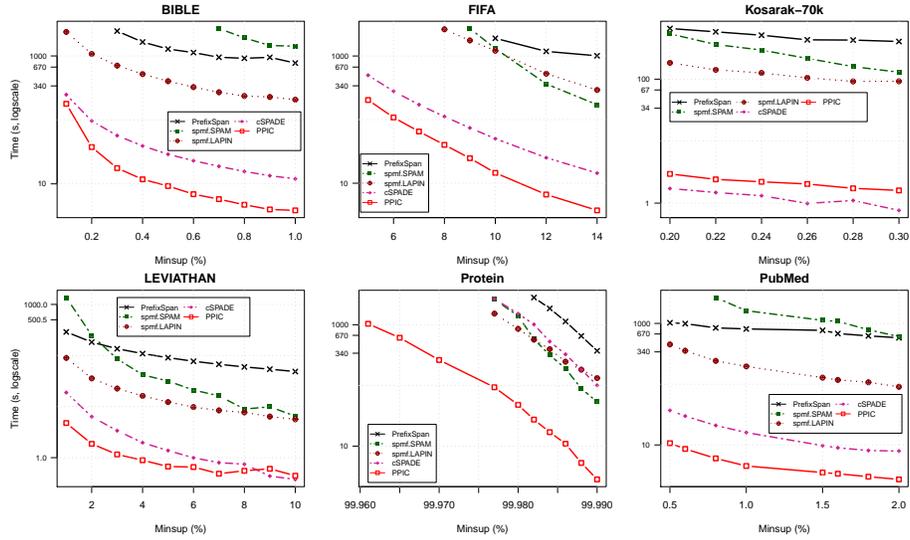}
\caption{CPU times for PPIC,PPDC,PPmixed and \textsc{cSPADE} for several $minsup$}
\label{fig:4}       
\end{figure*}

\paragraph{\textbf{Comparison with specialized algorithms.}} Our third experience is the comparison with  specialized algorithms. As we can see in the Fig.~\ref{fig:4}, we perform better on $84\%$ of the datasets. 
However, \textsc{cSpade} is still the most efficient for Kosarak.
In fact, Kosarak doesn't contain any symbol repetition in its sequences. So it is a bad case for prefix-projection-based algorithms which need to scan all the positions. On the contrary, with protein dataset (the sparse one) \textsc{cSpade} requires much more CPU time. The SPMF implementation of SPAM, PrefixSpan and LAPIN appears to be consistently slower than \textsc{cSpade} but there is no clear domination among these.

\begin{figure*}[!h]
  \includegraphics[width=\textwidth]{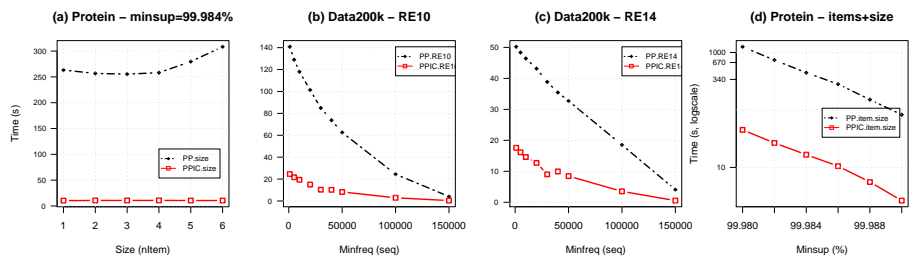}
\caption{Handling of different additional constraints  }
\label{fig:5} 
\end{figure*}

\paragraph{\textbf{Handling different additional constraints.}}  In order to illustrate the modularity of our approach we compare with a number of user-defined constraints that can be added as additional modules without changing the main propagator (Fig.~\ref{fig:5}). (a)
We compared PPIC and PP (unfortunately the \textsc{Gap-Seq} tool does not support a regular expression command-line argument) under various size constraints on the protein dataset with $minsup=99.984$.
(b,c) We also selected data200k adding a regular expression constraint  $RE10 = A*B(B|C)D*EF*(G|H)I*$ and $RE14 = A*(Q | BS*(B|C)) D* E (I|S)* (F|H) G* R$ \cite{trasarti2008sequence}. 
The last experiment reported on Fig.\ref{fig:5}d consists in combining size and symbols constraints on the protein dataset: only sequential patterns that contain VALINE and GLYCINE twice and ASPARATE and SERINE once are valid. 
PPIC under constraints still dominates PP.
 
\section{Conclusion}
\label{conclo}

This work improved the existing CP-based sequential pattern mining approaches \cite{negrevergne2015constraint,kemmar2015global} up to the point that it also surpasses specialized mining systems in terms of efficiency. 
To do so, we combined advanced ideas from the sequence mining literature, namely last-position lists \cite{lapin-spam}, as well as from the CP literature, namely memory-efficient data-structures for storing and restoring state during backtracking search. 
We introduced the \textsc{PrefixProjection-Inc} (PPIC), \textsc{PrefixProjection-Dec} (PPDC) and \textsc{PrefixProjection-mixed} (PPmixed) global constraints proposing different strategies to compute the projected frequencies: from scratch, by decreasing them, or a mix of both. 
These can be plugged in as modules in a CP solver.
These constraints are implemented in Scala and made available in the generic OscaR solver. 
Furthermore, the approach is compatible with a number of constraints including size and regular expression constraints. 
There are other constraints which change the subsequence relation and which would hence require hardcoding changes in the propagator (gap~\cite{kemmar2015global}, span, etc). We think many of our improvements can be applied to such settings as well.

Our work shows that generic CP solvers can indeed be used as framework to build scalable mining algorithms, not just for generic yet less scalable systems as was done for itemset mining~\cite{guns2011itemset}. Furthermore, advanced data-structures for backtracking search, such as trailing and reversible vectors, can be used in non-CP based mining algorithms too. We believe there is much more potential in such combinations of techniques from data mining and CP.


\bibliographystyle{splncs}       
{\footnotesize
\bibliography{biblio}}

\end{document}

%% file: fig1.tex
\begin{tikzpicture}

\tikzstyle{val}=[inner sep=1pt,minimum width=18pt,minimum height=18pt,anchor=north west]
\tikzstyle{valx}=[inner sep=1pt,minimum width=18pt,minimum height=15pt,anchor=north west]
   \tikzstyle{map}=[val]
   \tikzstyle{size}=[draw,thick]
   \tikzstyle{rng}=[<->,very thin,shorten <=1pt,shorten >=1pt]

\node[minimum width=18pt] (a) at (-18pt,0) {
		\begin{tabular}{ll}
		\noalign{\smallskip}\hline\noalign{\smallskip}
		$sid_1$ & $\langle ABCBC \rangle$   \\
		$sid_2$ & $\langle BABC \rangle$   \\
		$sid_3$ & $\langle AB \rangle$ \\
		$sid_4$ & $\langle BCD \rangle$   \\
		\noalign{\smallskip}\hline
		\end{tabular}
  };
  
  \node[minimum width=18pt] (b) at (72pt,0) {
		\begin{tabular}{ll}
		\noalign{\smallskip}\hline\noalign{\smallskip}
		$sid_1$ & $\langle BCBC \rangle$   \\
		$sid_2$ & $\langle BC \rangle$   \\
		$sid_3$ & $\langle B \rangle$ \\
		\noalign{\smallskip}\hline
		\end{tabular}
  };
  
 \node[minimum width=18pt] (c) at (162pt,0) {
		\begin{tabular}{ll}
		\noalign{\smallskip}\hline\noalign{\smallskip}
		$sid_1$ & $\langle CBC \rangle$   \\
		$sid_2$ & $\langle C \rangle$   \\
		$sid_3$ & $\langle  \rangle$ \\
		\noalign{\smallskip}\hline
		\end{tabular}
};

\node[minimum width=18pt] (d) at (232pt,0) {$\dots$};

\draw[->] (a) -- (b); 
\draw[->] (b) -- (c);
\draw[->] (c) -- (d);

\node[val] (e) at (20pt,15pt) {A};
\node[val] (f) at (110pt,15pt) {B};
\node[val] (g) at (200pt,15pt) {C};

\foreach[count=\xi from -1] \x in {0,1,2,3,4,5,6,7,8,9,10,11,12,13,14}
     {\expandafter\xdef\csname MAP-\x\endcsname{\xi}
      \node[valx,gray] (x) at (\xi*18pt,-68pt) {$\x$\strut};
      }

   \node[val] (deb) at (-46pt,-80pt) {$sids = $};
   \foreach[count=\xi from -1] \x in {1,2,3,4,1,2,3,1,2,3,.,.,.,.,.}
     {\expandafter\xdef\csname MAP-\x\endcsname{\xi}
      \node[val] (dom-\x) at (\xi*18pt,-80pt) {$\x$\strut};
      \draw[gray, thick,dotted] (\xi*18pt+18pt,-80pt) -- (\xi*18pt+18pt,-98pt) ;
      }
   \draw (-18pt,-80pt) rectangle +(\xi*18pt+36pt,-18pt);

   \node[val] (deb) at (-46pt,-100pt) {$poss = $};
   \foreach[count=\xi from -1] \x in {0,0,0,0,1,2,1,2,3,2,.,.,.,.,.}
     {\expandafter\xdef\csname MAP-\x\endcsname{\xi}
      \node[val] (dom-\x) at (\xi*18pt,-100pt) {$\x$\strut};
      \draw[gray, thick,dotted] (\xi*18pt+18pt,-100pt) -- (\xi*18pt+18pt,-118pt) ;
      }
   \draw (-18pt,-100pt) rectangle +(\xi*18pt+36pt,-18pt);
   
   \node[val] (c) at (-47pt,-40pt) {$(\phi=0,\varphi=4)$};
   \draw[blue, thick, dashed] (-18pt,-58pt) -- (-18pt,-121pt) ;
   \draw[blue, thick, dashed] (-47pt,-42pt) rectangle +(60pt,-16pt);
   
    \node[val] (c) at (25pt,-40pt) {$(\phi=4,\varphi=3)$};
   \draw[blue, thick, dashed] (54pt,-58pt) -- (54pt,-121pt) ;
   \draw[blue, thick, dashed] (25pt,-42pt) rectangle +(60pt,-16pt);
   
   \node[val] (c) at (90pt,-40pt) {$(\phi=7,\varphi=3)$};
   \draw[blue, thick, dashed] ( (108pt,-58pt) -- (108pt,-121pt) ;
   \draw[blue, thick, dashed] (90pt,-42pt) rectangle +(60pt,-16pt);
   
    \node[val] (c) at (154pt,-40pt) {$\dots$};
   \draw[blue, thick, dashed] ( (162pt,-58pt) -- (162pt,-121pt) ;
   

\end{tikzpicture}